\documentclass[journal=jpcafh,manuscript=article]{achemso}
\usepackage{longtable}
\usepackage{multirow}
\usepackage{amsmath}
\usepackage{subfigure}
\usepackage[usenames,dvipsnames]{xcolor}
\usepackage{soul}
\usepackage{bm}

\usepackage{xr}
\usepackage{amssymb}
\usepackage{siunitx}
\usepackage{threeparttable}
\usepackage{multicol} \usepackage{wrapfig} \usepackage{enumitem}
\usepackage{bm} \usepackage{outlines} 
\usepackage{booktabs}
\usepackage{graphicx}
\usepackage[normalem]{ulem}
\usepackage{hyperref}
\usepackage{lineno}
\usepackage[version=4]{mhchem}
\usepackage[section]{placeins}

\makeatletter
\newcommand*{\addFileDependency}[1]{
  \typeout{(#1)}
  \@addtofilelist{#1}
  \IfFileExists{#1}{}{\typeout{No file #1.}}
}
\makeatother

\newcommand*{\myexternaldocument}[1]{%
    \externaldocument{#1}%
    \addFileDependency{#1.tex}%
    \addFileDependency{#1.aux}%
}
\myexternaldocument{si}

\author{Raidel Martin Barrios, Abhirami Vijayakumar, Jingchun Wang}
\affiliation[University of Basel]{Department of Chemistry, University
  of Basel, Klingelbergstrasse 80, CH-4056 Basel, Switzerland.}

\author{Markus Meuwly} \affiliation[University of Basel]{Department of
  Chemistry, University of Basel, Klingelbergstrasse 80, CH-4056
  Basel, Switzerland.}\email{m.meuwly@unibas.ch}

\title{Towards Quantitative Reaction Dynamics of O$_3$}

\begin{document}
\date{\today}

\begin{abstract}
The reaction dynamics of O($^3$P) + O$_2(^3\Sigma_g^{-} )$ collisions
in the O$_{3}(^{1}A')$ electronic ground state is characterized on a
high-level MRCI+Q/aug-cc-pVQZ potential energy surface represented as
a reproducing kernel. For the atom exchange reactions involving the
$^{16} {\rm O}$ and $^{18}{\rm O}$ isotopes as the atomic collision
partner, associated with rates $k_6(T)$ and $k_8(T)$, respectively, a negative
temperature-dependence of $k(T)$, consistent with experiments was
found. The absolute rates typically underestimate measured rates by
$\sim 50$ \%, depending on the experiment considered. For the ratio
$R(T) = k_8(T) / k_6(T)$, the measured $T-$dependence was found,
including a cusp at lower temperatures. The differences between
experiments and computations are primarily due to neglect of
quantum effects, primarily zero-point effects. For the atomization reaction, leading to $3 {\rm
  O}(^3 {\rm P})$, the rates is lower by approximately one order of
magnitude compared with experiments, which is a clear improvement over
simulations using previous potential energy surfaces computed with
smaller basis sets. Nonadiabatic effects are deemed minor for the
atom exchange reactions.
\end{abstract}

\section{Introduction}
Quantitative simulations for small to medium-sized molecules have come
into reach over the past few years. One of the deciding factors was to
introduce machine learning-based techniques for representing
full-dimensional potential energy surfaces (PESs). Another driver was the
possibility to carry out electronic structure calculations at
sufficiently high levels of theory. Increasingly accurate simulations
are not only of interest when comparing measurements with observables
from computations. Quantitative gas phase reaction dynamics also
provides information about the validity of classical versus quantum
treatments of the nuclear dynamics, the relevance of including
particular quantum effects, the required level of electronic structure
theory for particular observables, or the accuracy with which ML-based
representations need to reproduce the underlying reference
calculations.\\

\noindent
The reaction dynamics of the O($^3$P) + O$_2(^3\Sigma_g^{-} )$ system
is of considerable interest in atmospheric chemistry and in rarefied
gas
flow.\cite{MM.sts:2019,MM.std:2022,MM.hyper:2020,Park_book,park1993review,panesi:2016}
This regime is encountered at the high altitudes relevant to
hypersonic flight.\cite{sarma:2000,walpot:2012,dsmc:2017} From a
chemical perspective, the processes occurring in hypersonic flight can
be viewed as a complement to ordinary burning whereby both lead to
chemical transformation of the starting material. While "burning" is
exothermic and releases chemical energy through fuel oxidation,
hypersonic vehicles experience intense aerodynamic heating due to
strong mechanical compression (shock waves with local $T \sim 10000$ K
or higher), which converts kinetic energy into internal energy. The
resulting high post-shock temperatures can dissociate atmospheric
N$_2$ and O$_2$, triggering high-temperature air chemistry to generate
specifically nitric oxide (NO) via the thermal (Zel’dovich)
mechanism,\cite{zeldovich:1946,bose:1996,MM.n2o:2020,singh:2018}
involving reactions between dissociated nitrogen and oxygen
species. \\
  
\noindent
In the wider context of the Zel’dovich (thermal NO production)
mechanism, the O+O$_2$ chemistry matters mainly because it controls
the availability of atomic oxygen, which is the key reactant in the
rate-limiting step N$_2$+O$\leftrightarrow$NO+O, followed by the fast
propagation step O$_2$+N$\Leftrightarrow$NO+O. In hot air (e.g.,
shock-heated or post-flame gases), NO formation is therefore highly
sensitive to the local O-atom pool. Any pathway that converts O into
more stable oxygen species reduces the driving reactant for the
N$_2$+O reaction and can suppress generation of thermal NO. On the
other hand, O-generating pathways can sustain formation of thermal
NO.\\

\noindent
In cooler or lower-temperature oxygen-rich regions (more typical of
upper-atmosphere/vacuum and some rarefied near-wall zones than of the
hottest shock layer), the relevant O$_2$+O sink is often the
termolecular association O$_2$+O+M$\rightarrow$O$_3$+M reaction, which
is collision-limited and can act as an O-atom reservoir/sink that
indirectly modulates the Zel’dovich chain through changing [O]
concentration. Importantly, the entire chemistry in hypersonic flight
is shock-induced and evolves under extreme non-equilibrium conditions,
with populations far from equilibrium with respect to occupation of
translational, rotational, vibrational, and electronic degrees of
freedom.\\

\noindent
A particular motivation for the present work is provided by the
observation that recent quasi-classical trajectory (QCT) simulations
on good-quality PESs underestimated measured dissociation rates for
the O($^3$P) + O$_2(^3\Sigma_g^{-} )$ by two orders of magnitude. One
of the debated ingredients for treating such processes is the
electronic degeneracy for which different values (ranging from 1/27 to
16/3) have been proposed in the literature. With a higher-level
treatment it should be possible to considerably narrow down this
interval. Another reason to consider the O($^3$P) +
O$_2(^3\Sigma_g^{-} )$ reaction system is the fact that two rather
different PESs yielded comparable $k(T)$ for the atom exchange and
dissociation reactions. This indicates that consistency with
experiments is not coincidental and improving one of the two provides
additional insights.\\

\noindent
The present work also considers isotopic effects on the atom exchange
reaction.\cite{fleurat:2003} This aspect is of particular relevance in
the wider context of isotopic fractionation in ozone with particular
emphasis on atmospheric chemistry.\cite{thiemens:1983} Ozone formation produces a
distinctive mass-independent fractionation (MIF) in oxygen isotopes
(an anomalous $\Delta^{17}{\rm O}$ signal) that cannot be explained by
ordinary mass-dependent kinetics. Hence, the MIF an important diagnostic of ozone chemistry and its coupling to broader oxidant
cycles. This “ozone anomaly” was first demonstrated in laboratory
ozone production and established that chemical reactions can generate
large, non-mass-dependent isotope effects in oxygen, fundamentally
changing how atmospheric isotope signals are
interpreted.\cite{thiemens:1983} Building on this, mechanistic work
showed that the anomaly is rooted in molecular-level dynamics of the
ozone-forming recombination reaction. Hence, ozone
isotopologues provide a sensitive probe of formation/stabilization
pathways under different conditions.\cite{babikov:2013-2} Crucially,
ozone’s $\Delta^{17}{\rm O}$ signature propagates into secondary
atmospheric products through ozone-involving oxidation (e.g., pathways
leading to nitrate).\\

\noindent
The present work is structured as follows. First, the methods are
presented, followed by the validation of the PES. Next, the results
from QCT simulations for the atom exchange and atomization reactions
are discussed. Finally, conclusions are drawn.\\

\section{Methods}

\subsection{Reactive potential energy surface}
The ground-state reactive PES of the system O$_{3}(^{1}A')$ was
constructed using high-level \emph{ab initio} electronic structure
calculations combined with a reproducing kernel Hilbert space (RKHS)
representation. All reference energies were computed at the
multireference configuration interaction level including the Davidson
correction
(MRCI+Q)\cite{langhoff1974configuration,werner1988efficient},
employing the augmented correlation-consistent polarized
quadruple-zeta basis set (aug-cc-pVQZ,
AVQZ).\cite{dunning1989gaussian}\\

\noindent
Electronic structure calculations were carried out using Jacobi
coordinates, where $R$ denotes the distance between the incoming atom
and the center of mass of the diatomic fragment, $r$ is the diatomic
bond length, and $\theta$ is the angle between the vectors
$\textbf{R}$ and $\textbf{r}$. Calculations were performed assuming
$C_{\rm S}$ symmetry using the \textsc{Molpro} suite of programs
\cite{werner2020molpro}. The reference data set was generated on a
three-dimensional grid spanning the physically relevant regions of
configuration space, including both bound and near-dissociation
geometries. The grid comprising $R \in [1.4,13.0]\,a_0$, $r \in
[1.1,4.16]\,a_0$, and $\theta \in [90.1^\circ,169.8^\circ]$ following
the same methodology of previous work.\cite{MM.o3:2025} The spacing
was non-uniform and refined in the interaction region, with denser
sampling at short intermolecular separations and near the equilibrium
bond length of the diatomic fragment. This results in a total of 2550
geometrically distinct and physically meaningful ground-state
configurations.\\

\noindent
To properly describe the electronic structure along the reactive
coordinates, state-averaged complete active space self-consistent
field CASSCF$(12,9)$ calculations were first carried out and used as
reference wave functions for the subsequent MRCI+Q computations. The
MRCI+Q approach was adopted instead of conventional MRCI in order to
improve the overall accuracy of the electronic structure calculations,
as the inclusion of the Davidson correction significantly reduces the
size-consistency error inherent to standard MRCI
\cite{rintelman2005multireference,shiozaki2011explicitly}. Multiple
electronic states were included in the state-averaging procedure to
ensure smooth behavior of the PES across different regions of
configuration space. In cases where convergence issues were
encountered or where reference points were missing, the data set was
completed using lower-dimensional RKHS interpolations prior to
constructing the full three-dimensional surface. The final
three-dimensional RKHS representation was generated using the
\textsc{kernel-toolkit} \cite{MM.rkhs:2017}.\\

\noindent
The global reactive PES was constructed by combining three
single-channel PESs corresponding to the three possible atom-diatom
arrangements using a permutationally invariant mixing scheme. The
total potential was written as
\begin{equation}
V(\vec{r}) = \sum_{i=1}^{3} w_i(r_{i}) V_i(\vec{r}),
\end{equation}
where $\vec{r} = (r_{AB}, r_{AC}, r_{BC})$ denotes the set of
interatomic distances. The channel weights $w_i$ were defined using an
exponential switching function,
\begin{equation}
w_i(r_{i}) = \frac{\exp\left[-(r_i/\rho)^2\right]}{\sum_{j=1}^{3} \exp\left[-(r_j/\rho)^2\right]},
\label{eq:mix}
\end{equation}
with a single optimized switching parameter $\rho$. The value of
$\rho = 1.63$ $a_{0}$ was determined through a grid-based optimization
using an independent mixing data set specifically designed to sample
the regions where multiple channels overlap. The mixing data set was
constructed on a grid defined by angular configurations $\theta =
{30^\circ, 60^\circ, 90^\circ, 120^\circ, 150^\circ}$, while the
interatomic distances $r_{AB}$ and $r_{BC}$ were sampled identically
in the range $2.30$–$4.00,a_0$, using a non-uniform spacing refined in
the channel crossing region.\\

\noindent
The quality of the PES was assessed by evaluating the root-mean-square
deviation (RMSD) and correlation coefficient $r^2$ for on-grid,
mixing, and off-grid validation data sets. The off-grid validation
geometries were generated using angular values $\theta = {20^\circ,
  40^\circ, 80^\circ, 130^\circ, 160^\circ}$, with both interatomic
distances $r_{\rm AB}$ and $r_{\rm BC}$ spanning the interval $1.9$–$5.1,a_0$.
These configurations were excluded from both the construction of the
single-channel PESs and the optimization of the mixing parameters in
the channel-crossing regions. The resulting PES accurately reproduces
the reference MRCI+Q/AVQZ energies over a wide energy range, including
configurations well above the dissociation limit.\\

\subsection{Quasi-classical trajectory simulations}
Quasi-classical trajectory simulations were performed on the
constructed RKHS-PES to investigate the reaction dynamics and compute
thermal rate coefficients. Initial conditions were sampled
semiclassically from Boltzmann distributions of translational energy
$E_{\textrm{trans}}$, impact parameter $b$, total angular momentum
$J$, and rovibrational states ($v$,$j$) of the diatomic molecule. To
ensure statistical convergence, $5 \times 10^{6}$ independent
quasi-classical trajectory simulations were performed at each
temperature. The trajectories were propagated until any interatomic
distance exceeded a cutoff of $20\,a_{0}$, indicating that the
fragments were fully separated (atomization), or until the maximum
simulation time of $75$ ps was reached. The reaction outcomes were
then identified using geometrical criteria based on the interatomic
distances. Atom exchange was assigned when one pair of atoms formed a
diatom ($r_{ij}<8\,a_{0}$) while the other two interatomic distances
exceeded $16\,a_{0}$, indicating the formation of a new diatomic bond
and the separation of the third atom.\\

\noindent
Thermal rate coefficients at a given temperature $T$ were computed
according to
\begin{equation}
k(T) = g_e(T)\,\sqrt{\frac{8 k_B T}{\pi \mu}}\,\pi b_{\mathrm{max}}^2
\frac{N_r}{N_{\mathrm{tot}}},
\end{equation}
where $\mu$ is the reduced mass, $b_{\mathrm{max}}$ is the maximum
impact parameter, $N_r$ is the number of reactive trajectories, and
$N_{\mathrm{tot}}$ is the total number of trajectories, and
$k_{\textrm{B}}$ is the Boltzmann constant. The factor $g_e(T)$
accounts for the electronic degeneracy of the reaction channel under
consideration. For the atom exchange process, the
temperature-dependent electronic degeneracy factor
was\cite{fleurat:2003-2,guo:2013,gross1997isotope}
\begin{equation}
	g_{e}^{\mathrm{exch}}(T) = 1/\{3\left[5 + 3\,\exp(-277.6/T) + \exp(-325.9/T)\right]\}.
\end{equation}
In contrast, for the full dissociation reaction and in the absence of
nonadiabatic effects, a temperature-independent electronic degeneracy
of $g_{e}^{\mathrm{diss}} = 1/27$ was used here, also following
earlier work.\cite{fleurat:2003-2} It is worth noting that alternative
temperature-independent electronic degeneracy factors have been
adopted in previous studies, most notably a value of
$g_{e}^{\mathrm{diss}} = 16/3$. This choice, however, was originally
introduced for the Ar + O$_2$ dissociation system and later
transferred to the present context. The two values
$g_{e}^{\mathrm{diss}} = 16/3$ and $g_{e}^{\mathrm{diss}} = 1/27$
should therefore be regarded as approximate upper and lower limits,
respectively. As a consequence, the dissociation rate
$k_{\mathrm{diss}}(T)$ reported in this work, which is based on
$g_{e}^{\mathrm{diss}} = 1/27$, represents a lower bound to the
corresponding temperature-dependent rate. Statistical uncertainties
were estimated using bootstrap resampling.\cite{bootstrapping:1993} For each
temperature, the full trajectory set was divided into multiple
batches, and random resampling was used to obtain standard deviations
for the computed rates.\\

\section{Results}

\subsection{Validation of the AVQZ-RKHS-PES}

\begin{figure}[h!]
  \centering
  \includegraphics[width=0.9\textwidth]{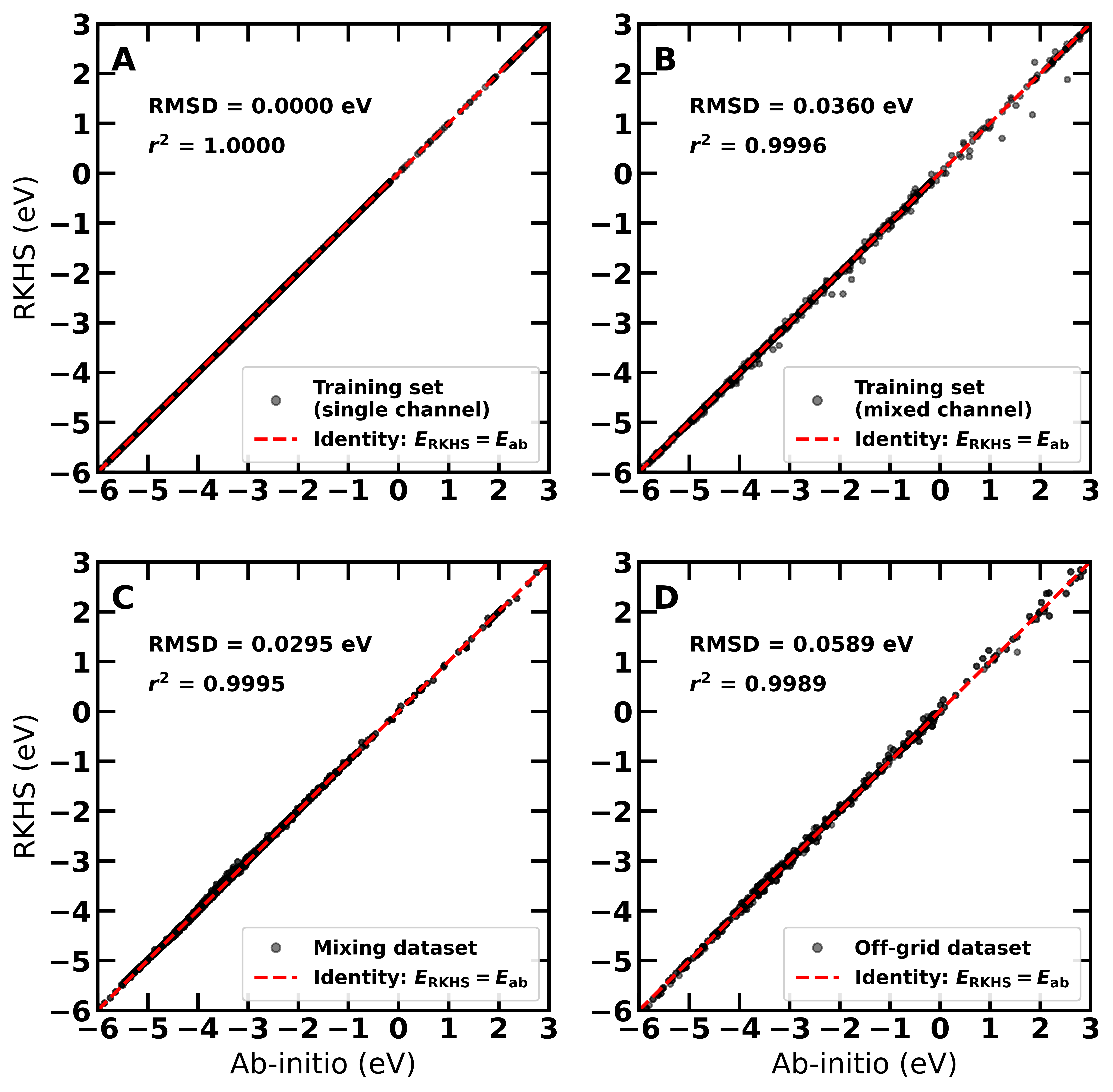}
  \caption{Performance of the RKHS-represented reactive,
    three-dimensional PES for the [OOO] system with the zero of energy
    at the O($^3$P)+O($^3$P)+O($^3$P) asymptote. (A) Correlation
    between single-channel RKHS representation and reference energies
    for on-grid training data. (B) Correlation between 3D mixed RKHS
    representation and reference energies for the on-grid training
    data. (C) Performance of the 3D mixed RKHS representation on the
    mixing region. (D) Performance of the 3D mixed RKHS representation
    for 1188 off-grid points not used in constructing the PES. The
    RMSD between the RKHS representation and the reference data and
    the corresponding $r^{2}$ are given in each panel.}
  \label{fig:validation}
\end{figure}

Figure \ref{fig:validation} reports statistical performances of the
non-reactive and reactive potential energy surfaces. The
single-channel PES (Figure \ref{fig:validation}A) featured a
representation error of ${\rm RMSD} < 10^{-5}$ eV (0.0002 kcal/mol)
and $r^2 = 1.0$ across 9 eV. Such a performance is on par with
previous work.\cite{MM.o3:2025,MM.n3:2024} Panel B reports the
performance of the mixed surface, see Eq. \ref{eq:mix}, on the
reference data set. The RMSD$(E)$ increases to 0.0360 eV with $R^2 =
0.9996$. In order to further refine the mixed PES around the crossing
regions, the mixing parameter was refined by fitting to the mixing
data set which features an RMSD$(E)$ of 0.0295 eV and associated $R^2
= 0.9995$, see Figure \ref{fig:validation}C.\\

\noindent
Finally, the performance of the mixed, 3-channel PES was assessed on a
test set in which all geometries are ``off-grid" with respect to all
available configurations used to construct the single- and 3-channel
PESs. Figure \ref{fig:validation}D underscores the high quality of the
reactive PES for the [OOO] system with ${\rm RMSD}(E) = 0.0589$ eV
(1.35 kcal/mol) across an energy range of 9 eV (210
kcal/mol). Evidently, the performance is considerably better for
energies below the O($^3$P)+O($^3$P)+O($^3$P) asymptote. As a
comparison, the performance of a recent PIP-based PES across 200
kcal/mol was ${\rm RMSD}(E) \sim 4.5$ kcal/mol.\cite{varga:2017}\\

\begin{figure}[H]
  \centering
  \includegraphics[width=0.98\textwidth]{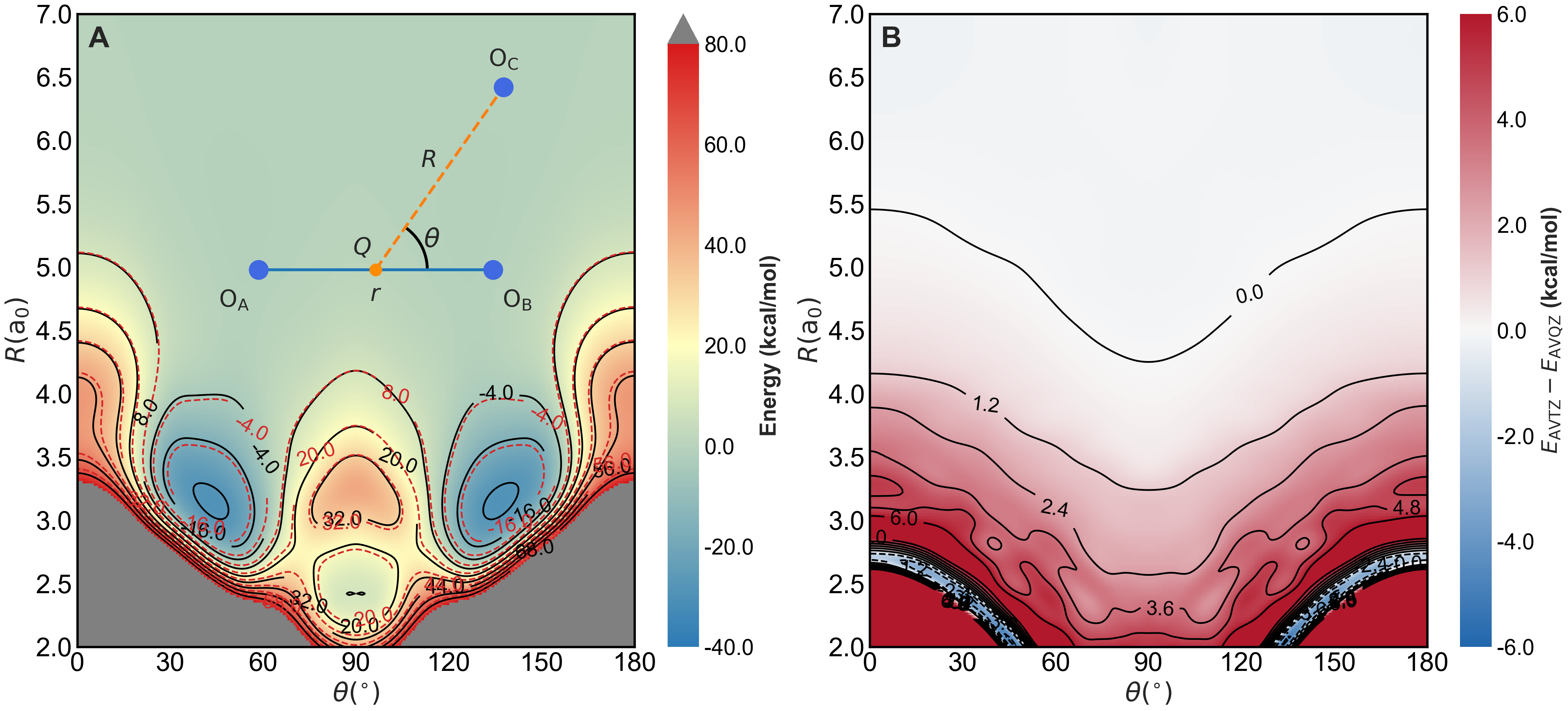}
  \caption{Panel A: Two-dimensional potential energy surface (PES) for
    a fixed internuclear distance $r = 2.46$~a.u. as a function of the
    Jacobi coordinates $(R,\theta)$, computed at the AVQZ level.  The
    color scale represents the energy in kcal/mol.  Solid black
    contour lines correspond to AVQZ energies, while dashed red
    contours indicate AVTZ results for comparison. The inset
    illustrates the definition of the Jacobi coordinates: $r$ is the
    diatomic bond length between ${\rm O}_{\rm A}$ and ${\rm O}_{\rm
      B}$, $R$ is the distance from the diatomic center of mass $Q$ to
    atom ${\rm O}_{\rm C}$, and $\theta$ is the angle between
    $\vec{r}$ and $\vec{R}$.  Panel B: Energy difference surface
    $E_{\mathrm{AVTZ}} - E_{\mathrm{AVQZ}}$ as a function of
    $(R,\theta)$.  The color map highlights regions where AVTZ
    overestimates (red) or underestimates (blue) the AVQZ energies.
    Contour lines are drawn at uniform intervals. The minimum energy
    in the reactant well region is found to be $E^{\mathrm{\min}}_{\rm
      AVQZ} = -29.6$ kcal/mol and $E^{\mathrm{\min}}_{\rm AVTZ} =
    -26.4$ kcal/mol within the selected $(R,\theta)$ window.}
\label{fig:difference}
\end{figure}

\noindent
Next, the overall shapes of the AVTZ- and AVQZ-RKHS potential energy
surfaces was compared. The AVTZ-RKHS surface was previously validated
from comprehensive QCT simulations vis-a-vis thermal rates $k(T)$ for
the $T-$dependence of the atom-exchange and the atomization
reaction.\cite{MM.o3:2025} Figure \ref{fig:difference} reports the
shapes $V(R,\theta)$ of the two PESs for given O--O separation. As
Figure \ref{fig:difference}A shows, the isocontours for the two PESs
closely follow each other and the total interaction energy using the
AVQZ basis set is consistently lower than for the AVTZ basis set. As
an example, the stabilization of the O$_3$ structure $(R = 3.14~{\rm
  a}_0, \theta = 43.6^\circ)$ between the AVTZ- and AVQZ-RKHS PESs
increases from $-26.4$ kcal/mol to $-29.6$ kcal/mol. This indicates
that for given collision energy the thermal rates on the higher-level
AVQZ-PES are expected to decrease compared with those from QCT
simulation using the AVTZ-PES.\\

\subsection{The Atom Exchange and Isotopic Substitution Reactions}
First, QCT simulations using the AVQZ-PES were carried out to
characterize the atom-exchange reaction for the natural and the heavy
isotope of the incoming oxygen atom. This reaction is characterized by
the exchange of the incoming oxygen atom O$_{\rm A}$ by one of the
oxygen atoms O$_{\rm B}$ or O$_{\rm C}$ forming the O$_2$ molecules:
${\rm O_A}+{\rm O_B}{\rm O_C} \rightarrow {\rm O_B}+{\rm O_A}{\rm
  O_C}$ or $\rightarrow {\rm O_C}+{\rm O_A}{\rm O_B}$. In addition to
${\rm ^{16}O_A}+{\rm ^{16}O_B}{\rm ^{16}O_C}$, simulations were also
carried out for isotopic variants to determine the rates $k_6(T)$ and
$k_8(T)$, respectively.\cite{fleurat:2003} Here, $k_6(T)$ and $k_8(T)$
refer to the $^{16} {\rm O}$ and $^{18}{\rm O}$ isotopes as the
collision partner of ${\rm ^{18}O}{\rm ^{18}O}$ and ${\rm ^{16}O}{\rm
  ^{16}O}$, respectively.\\

\begin{figure}[H]
  \centering
  \includegraphics[width=0.9\textwidth]{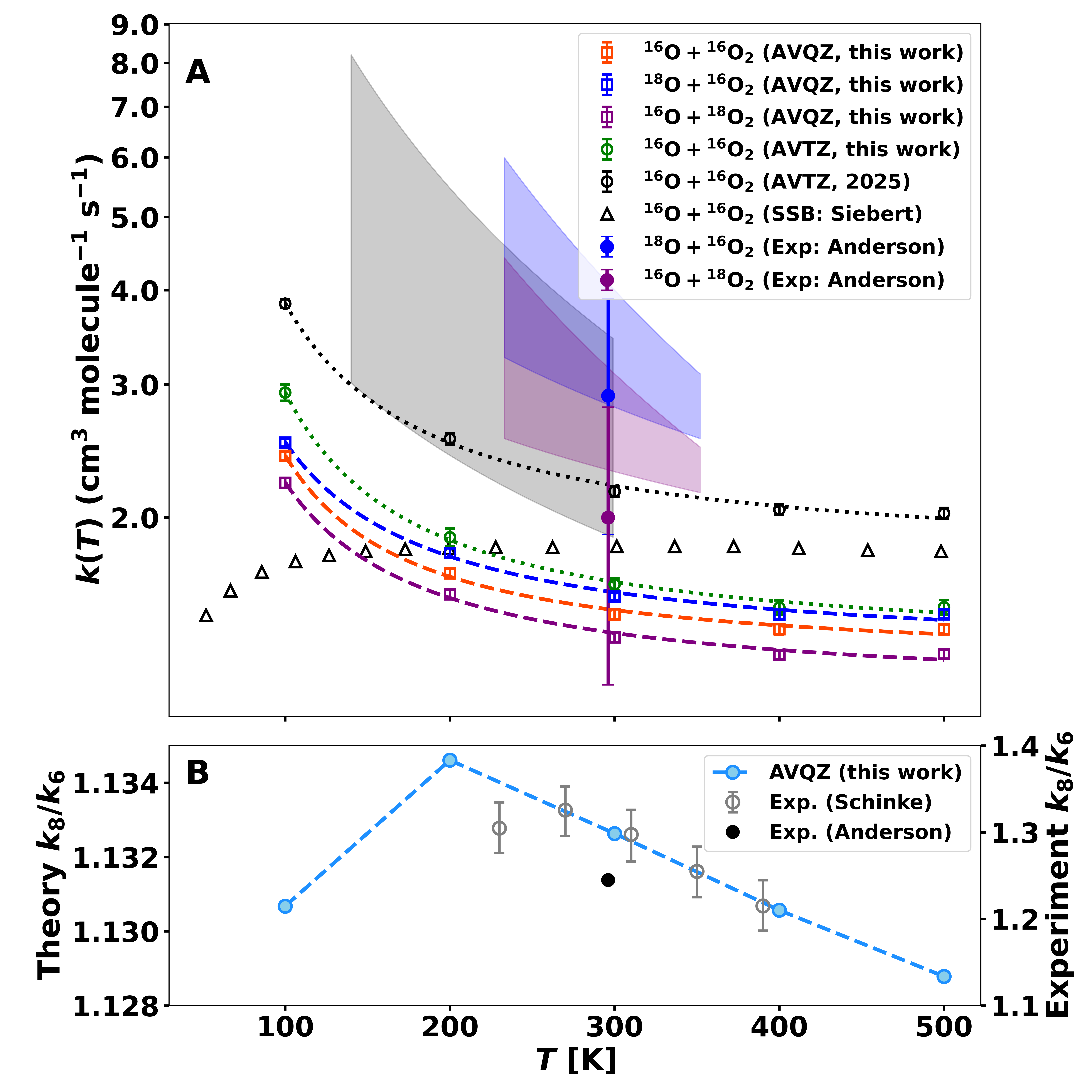}
\caption{Temperature dependence of the thermal rate coefficients
  $k(T)$ for the natural and heavy isotope of the incoming oxygen
  atom. Panel A: Results for $k(T)$ obtained from the RKHS-based PESs
  using the AVTZ and AVQZ basis sets and for oxygen isotopic
  substitutions as indicated in the legend. Earlier wave packet
  simulations using the SSB PES (black triangles) illustrate the
  negative $T-$dependence a number of previous studies
  reported.\cite{siebert2001spectroscopy} The measured
  rates\cite{anderson:1997} (blue and violet filled circles) are shown
  together with the reported uncertainties. Other measured rates are
  represented by the shaded uncertainty regions: blue and
  violet\cite{fleurat:2003} for $^{18}\mathrm{O} + ^{16}\mathrm{O}_2$
  and $^{16}\mathrm{O} + ^{18}\mathrm{O}_2$, respectively, while the
  gray\cite{wiegell1997temperature} region applies to both. Panel B:
  $T-$dependence of the isotopic rate ratio $R(T) =
  k_8(T)/k_6(T)$. Theoretical predictions obtained with the AVQZ
  RKHS-PES (blue symbols) are compared with
  measurements\cite{fleurat:2003,anderson:1997} (grey and black
  symbols).}
  \label{fig:rate_exch}
\end{figure}

\noindent
Using the new AVQZ-RKHS PES for the $^{16}{\rm O} + ^{16}{\rm
  O}^{16}{\rm O}$ exchange reaction (orange), the expected negative
$T-$dependence was found. In verifying the previous results using the
AVTZ-RKHS PES it was realized that an erroneous reduced mass was
employed in the earlier study.\cite{MM.o3:2025} The previous (black)
and corrected (green) exchange rates using the AVTZ-RKHS PES are also
reported in Figure \ref{fig:rate_exch}A. The $T-$dependence remains
unchanged but $k(T)$ decreases by $\sim 25$ \% using the correct
reduced mass. Finally, results for the systems for which measurements
have been carried out - $^{18}{\rm O} + ^{16}{\rm O}^{16}{\rm O}$
(blue) and $^{16}{\rm O} + ^{18}{\rm O}^{18}{\rm O}$ (violet) - are
also shown in Figure \ref{fig:rate_exch}B. With increasing mass of the
collision system the thermal rates decrease by a few percent. This is
consistent with two earlier experimental
studies\cite{fleurat:2003,anderson:1997} but at variance with a third
one.\cite{wiegell1997temperature} When comparing absolute values for
$k(T)$ it should be noted that the QCT simulations do not include zero
point vibrational energy which has been identified as an ingredient
required for quantitative comparisons between simulations and
experiment.\cite{fleurat:2003}\\

\noindent
The shaded black and colored regions in Figure \ref{fig:rate_exch}A,
show the reported experimental uncertainty ranges from the literature
data.\cite{wiegell1997temperature,schinke:2003} and values at 296 K
from another experimental study.\cite{anderson:1997} The $T-$dependent
experiments feature $k(T)$ that decreases with increasing
temperature. This observation and the fact that several simulation
studies reported a positive $T-$dependence has lead to intense
discussions regarding the overall shape of the PES. Specifically, a
so-called ``reef'' along the minimum energy path for the atomic oxygen
approaching the O$_2$ collision partner at long range had been
proposed to lead to the disagreement between measurements and
simulations.\cite{fleurat:2003,babikov:2003,ayouz:2013,Dawes:2013,Li:2014,tyuterev:2014}
Recently, however, it has been established\cite{MM.o3:2025} from QCT
simulations using two different high-level reactive and global PESs
(at MRCI+Q\cite{MM.o3:2025} and CASPT2\cite{varga:2017} levels) for
the [OOO] collision system that the "reef" is an integral feature of
the PESs and not an artifact and leads to the $T-$dependence
consistent with experiments. Furthermore, the "reef" has been known
for some time from electronic structure calculations at different
levels of theory.\cite{pack:2002,schinke:2004,holka:2010}\\

\noindent
Figure \ref{fig:rate_exch}B reports the $T-$dependence of the isotopic
rate ratio $R(T) = k_8(T)/k_6(T)$, which provides another sensitive
probe of the PESs and the underlying dynamics. The ratio $R(T)$ has
also been measured and found to have a cusp-like structure with a
maximum around 300 K (grey symbols in panel B).\cite{fleurat:2003} The
present simulations feature this cusp and the slope towards higher
temperatures but the magnitude is too low compared with the
experiments. Another experiment reported $R(T = 296{\rm K}) = 1.245$,
which is also indicated in the Figure (black).\cite{anderson:1997} As
already mentioned, including zero point energy differentially
influences the values of $k_6(T)$ and $k_8(T)$ and therefore affects
$R(T)$ as well. However, the overall behaviour of $R(T)$ is not
expected to change dramatically.\\

\subsection{The Atomization Reaction}
Next, the atomization reaction O($^3$P) + O$_2(^3\Sigma_g^{-} )$
$\rightarrow$ 3O($^3$P) was considered. For this process, measurements
to compare with are also
available.\cite{byron:1959,shatalov:1973}\noindent The QCT results
using the AVQZ-PES (dark red) correctly capture the $T-$dependence of
the measured rates (open circles and dashed lines). Using $g_{\rm
  e}^{\rm diss} = 1/27$ the absolute magnitude of the measured rates
is underestimated by $\sim$one order of magnitude. This compares with
a difference of $\sim$two orders of magnitude from simulations using
triple-zeta-quality PESs (red and blue symbols) from previous
work,\cite{MM.o3:2025} which also employed $g_{\rm e}^{\rm diss} =
1/27$. Hence, the higher-level PES clearly improves the agreement with
experiment but still underestimates the magnitude of the dissociation
rate.\\

\begin{figure}[h!]
  \centering \includegraphics[width=0.9\textwidth]{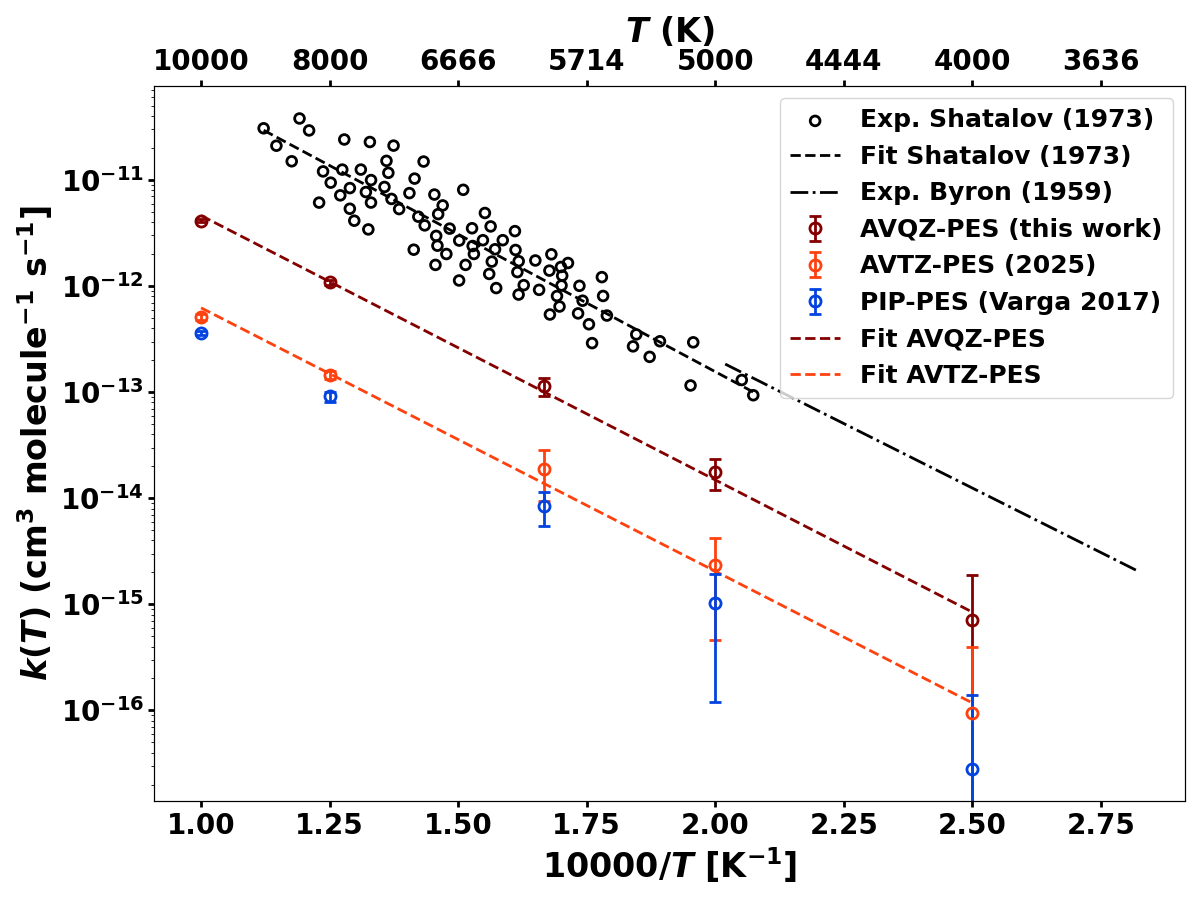}
  \caption{Thermal rate coefficients $k(T)$ for the dissociation
    reaction using $g_{\mathrm{diss}} = 1/27$.  The rate constants are
    shown as a function of inverse temperature ($10000/T$) on the
    lower horizontal axis, while the corresponding temperature scale
    (in K) is displayed on the upper axis.  Simulation results
    obtained with different potential energy surfaces are shown as
    colored symbols, including statistical error bars estimated via
    bootstrapping.  Results based on the RKHS-PES constructed with the
    AVTZ\cite{MM.o3:2025} and AVQZ (present work) basis sets are shown
    as red open circles and dark red symbols, respectively, while
    those using the PIP-PES\cite{MM.o3:2025,varga:2017} are the blue
    symbols. To visually emphasize the temperature dependence, the
    AVTZ and AVQZ simulation results were fitted and are represented
    by dashed lines in their respective colors. Experimental
    measurements\cite{byron:1959,shatalov:1973} are shown as black
    open circles.  A linear regression of the experimental data is
    indicated by the dashed black line.}
  \label{fig:rate_diss}
\end{figure}

\noindent
The improvement demonstrates the sensitivity of the dissociation
dynamics to the quantitative accuracy of the underlying potential
energy surface.  The use of the larger aug-cc-pVQZ (AVQZ) basis set
provides a more complete treatment of electron correlation and a more
accurate representation of the interaction energies, which directly
impacts on the O$_2$ dissociation energy, as already reported
previously.\cite{MM.o3:2025} Consequently, the AVQZ-based PES yields
rate coefficients that are in substantially better agreement with the
experimental data\cite{byron:1959,shatalov:1973} than those obtained
with the AVTZ-based surface. This emphasizes the importance of
basis-set convergence for achieving quantitatively reliable reaction
dynamics simulations.\\

\section{Discussion and Conclusions}
The present work reports thermal rates for the atom exchange and
dissociation reactions for the O($^3$P) + O$_2(^3\Sigma_g^{-} )$
system using a high-level MRCI+Q/AVQZ PES represented as a RKHS. For
the atom exchange reaction the $T-$dependence is consistent with the
measurements whereas the magnitude of the rates is lower by a factor
of $\sim 2$. For the dissociation reaction the $T-$dependence of the
experiments is correctly captured whereas the magnitude of the rate is
lower by about one order or magnitude. Notably, the ratio $R(T) =
k_8(T) / k_6(T)$ which compares the rates for the $^{18}{\rm
  O}+^{16}{\rm O}_2$ exchange reaction with that for $^{16}{\rm
  O}+^{18}{\rm O}_2$ mirrors that reported from
measurements:\cite{fleurat:2003} starting from low-$T$, $R(T)$
increases, reaches a cusp and then decreases monotonically towards
higher $T$. However, the temperature at which the cusp appears is
lower by several 10 K and the magnitude of $R(T)$ is clearly
underestimated.\\

\begin{figure}[H]
  \centering
  \includegraphics[width=0.98\textwidth]{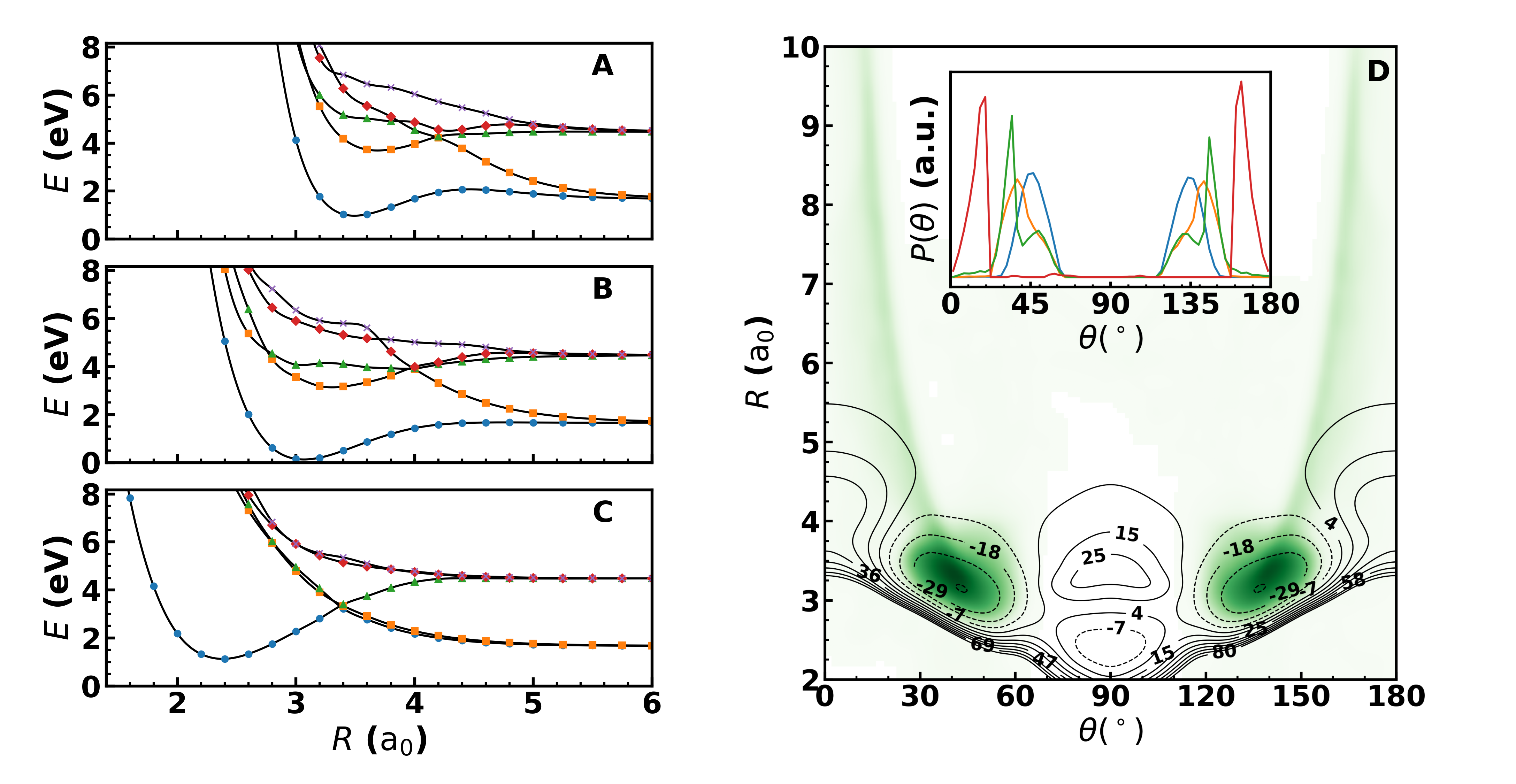}
  \caption{Potential energy curves of the five lowest $^1{\rm A}'$
    electronic states ($1\,^1{\rm A}'$ (blue), $2\,^1{\rm A}'$
    (orange), $3\,^1{\rm A}'$ (green), $4\,^1{\rm A}'$ (red), and
    $5\,^1{\rm A}'$ (purple)) as functions of the dissociation
    coordinate $R$ for fixed angles $\theta = 156.6^\circ$, $\theta =
    129.9^\circ$ and $\theta = 90.1^\circ$ (panels A to C). The O--O
    bond distance was $r = 2.65\,a_0$.  The zero of energy is the
    minimum of the ground-state surface (panel B).  In A to C, colored
    symbols denote adiabatic \emph{ab initio} MRCI energies, while
    solid black lines represent the corresponding diabatic potential
    energy curves. Panel D: Projection of the distribution
    $P(R,\theta)$ (green) from 1500 trajectories run at 100 K, leading
    to atom exchange onto the adiabatic ground-state potential energy
    surface $V(R,\theta; r = 2.65\,a_0)$. Inset: Angular distribution
    $P(\theta)$ for $R = 3.0\,a_0$ (blue), $R = 3.5\,a_0$ (orange), $R
    = 4.0\,a_0$ (green), and $R = 7$ a$_0$ (red). Note the
    near-symmetry of $P(\theta)$ with respect to $\theta = 90^\circ$.}
  \label{fig:cross}
\end{figure}

\noindent
It is also of interest to consider the possible role of crossings
between the singlet PESs. For this, the energies of the lowest 4
excited singlet states ($2\,^1{\rm A}'$ to $5\,^1{\rm A}'$) were
evaluated on the same grid as the $1\,^1{\rm A}'$ ground state. Up to
this point, the energies of the excited states were not represented as
a RKHS. This is outside the main scope of the present work but will
follow similar lines of previous work on the C+O$_2$
reaction.\cite{MM.co2:2021} It has already been reported that
non-adiabatic effects are unlikely to be the reason for disagreements
between simulations and measurements for the atom-exchange
reaction.\cite{schinke:2003,tyuterev:2018} In order to assess whether
using the larger AVQZ basis set leads to the same conclusion, part of
the relevant potential energy surfaces are considered in Figure
\ref{fig:cross}.\\

\noindent
One-dimensional cuts for $r_{\rm OO} = 2.65$ a$_0$ and three angles
$\theta$ are considered. Here, the zero of energy is the minimum of
the ground-state $1\,^1{\rm A}'$ surface (Figure \ref{fig:cross}B,
blue symbols). For structures around the minimum ($\theta =
156.6^\circ$ and $\theta = 129.9^\circ$; panels A and B) no crossings
between the ground and the excited state PESs are found. However,
distortion to $\theta = 90^\circ$ (panel C) leads to a crossing of the
$2\,^1{\rm A}'$ and $3\,^1{\rm A}'$ states with the $1\,^1{\rm A}'$
ground state at $R = 3.36$ a$_0$. Hence, in this region the exchange
reaction O($^3$P) + O$_2(^3\Sigma_g^{-} )$ asymptotically may lead to
excited state products ${\rm O} ^1{\rm D} + {\rm O}_2(a ^1\Delta{\rm
  g})$ if the trajectory follows the diabatic PES. The shapes and
relative energetics of the excited state surfaces are consistent with
earlier work.\cite{schinke:2005}\\

\noindent
Figure \ref{fig:cross}D illustrates the probability distribution
$P(R,\theta)$ of trajectories run at 100 K on the $1\,^1{\rm A}'$ PES
and leading to atom exchange on their way to products, projected onto
the adiabatic ground state PES $V(R,\theta;r = 2.65\,{\rm a}_0)$. For
regions in $(R,\theta)-$space for which crossings between the
$1\,^1{\rm A}'$ and the $2\,^1{\rm A}'$ / $3\,^1{\rm A}'$ surfaces
occur, the trajectories can potentially change electronic states due
to strong coupling. Instead of the ground state products O($^3$P) +
O$_2(^3\Sigma_g^{-} )$ after atom exchange, electronically excited
products will be formed. Such processes will reduce the rate for the
particular process considered, which is ``atom exchange'' in the
present case. However, the $(R,\theta)-$configurations sampled after
atom exchange has taken place demonstrates that the region in which
the $1\,^1{\rm A}'$ and the $2\,^1{\rm A}'$ / $3\,^1{\rm A}'$ PES
cross ($\theta \sim 90^\circ$) is not visited. This is most clearly
seen in the inset in Figure \ref{fig:cross}D. Hence, the present QCT
simulations indicate that nonadiabatic effects play a minor role for
the atom exchange reaction at $T$ up to 500 K which is consistent with
earlier findings.\cite{schinke:2003}\\

\noindent
Underestimating $k(T)$ for the exchange reaction is primarily a
consequence of neglecting zero-point energy (ZPE) in the QCT
simulations.\cite{fleurat:2003,tyuterev:2018} Hence, the present
finding is reassuring. Also, the rates from using the
RKHS/AVTZ\cite{MM.o3:2025} and RKHS/AVQZ PESs are consistent with one
another. Accounting for ZPE will also change $R(T)$. Finally,
nonadiabatic processes are unlikely to contribute significantly for
the atom exchange reaction because the crossing region between the
$1\,^1{\rm A}'$ ground state and the two lowest excited singlet states
($2\,^1{\rm A}'$ / $3\,^1{\rm A}'$) is not sampled for the
trajectories leaving the strongly interacting region after atom
exchange has taken place, see Figure \ref{fig:cross}D.\\

\noindent
As to $k(T)$ for the atomization reaction, it is found that the
agreement with experiment improves by one order of magnitude compared
with simulations based on the RKHS/AVQZ PES, using $g_e = 1/27$ as in
previous work\cite{fleurat:2003-2,MM.o3:2025} excluding nonadiabatic
effects. On the other hand, it should be noted that the ``correct''
electronic degeneracy is not known: $T-$independent values ranging
from $g_e = 16/3$ (borrowed from the Ar + O$_2$ dissociation
reaction\cite{nikitin:book}) down to $g_e = 1/27$ have been used in
the
literature.\cite{boyd:2016,mankodi:2017,cheng:2022,fleurat:2003-2,MM.o3:2025}
Both, the magnitude and assuming a $T-$independent value for $g_e$ are
idealizations.\\

\noindent
In conclusion, the RKHS/AVQZ PES represents a clear improvement in QCT
simulations for the atomization rate and corroborates the findings of
a negative $T-$dependence for the atom-exchange reaction. Notably,
the shape and $T-$dependence for $R(T)$ is consistent with
experiments. Studies as the one presented here demonstrate that
atomistic simulations for chemical reactions have reached a mature
state in which reasons for remaining disagreement between results from
computations and measurements can be further narrowed-down in order to
reach a more profound understanding of the processes in question.\\

\section*{Data Availability}
The codes and data for the present study are available from
\url{https://github.com/MMunibas/o3-avqz} upon publication.

\section*{Acknowledgment}
The authors gratefully acknowledge financial support from the Swiss
National Science Foundation through grants $200020\_219779$ (MM),
$200021\_215088$ (MM), and the University of Basel (MM). This article
is also based upon work within COST Action COSY CA21101, supported by
COST (European Cooperation in Science and Technology) (to MM).\\

\bibliography{bib}
\end{document}